\documentclass{revtex4}

\newcommand{\beq}{\begin{equation}}
\newcommand{\eeq}{\end{equation}}
\newcommand{\bea}{\begin{eqnarray}}
\newcommand{\eea}{\end{eqnarray}}

\usepackage{graphicx}
\begin{document}



\title{Tidal Waves Ð 
a non-adiabatic microscopic description
of the yrast states in near-spherical nuclei 
}

\author{S. Frauendorf, Y. Gu, and J. Sun}

\address{Department of Physics, University of Notre Dame, Notre Dame, IN 46556, USA\\
sfrauend@nd.edu.edu}


\begin{abstract}
The yrast states of  nuclei that are spherical or weakly deformed 
in their ground states are described as 
quadrupole waves running over the nuclear surface, which we  call  "tidal waves".
The energies and E2 transition probabilities of the yrast
states in  nuclides with $Z$= 44, 46, 48 and
$N=56, ~58,~...,~ 66$ are 
calculated by means of the cranking model in a microscopic way.
The nonlinear response of 
the nucleonic orbitals results in a strong coupling between shape and single particle 
degrees of freedom.
\end{abstract}
\maketitle
Nuclei are conventionally classified as as "rotors" and "vibrators":
Rotors have a fixed shape that goes round; vibrators execute
pulsating vibrations.  However this familiar view is incomplete, because 
the quadrupole surface oscillations are five dimensional. 
In this work, we make use of the fact that the yrast states of vibrational and transitional nuclei
have the very simple structure of a running surface waves, which we suggest calling "tidal waves".
Since they have a constant deformation in the co-rotating system of coordinates, one can
calculate their properties 
microscopically by means of the cranked mean field theory. A first report of the calculations was given in 
\cite{arXiv}, which used a less accurate interpolation. 
\begin{figure}[ht]
\begin{minipage}[b]{0.47\linewidth}
\centering
\includegraphics[width=\columnwidth]{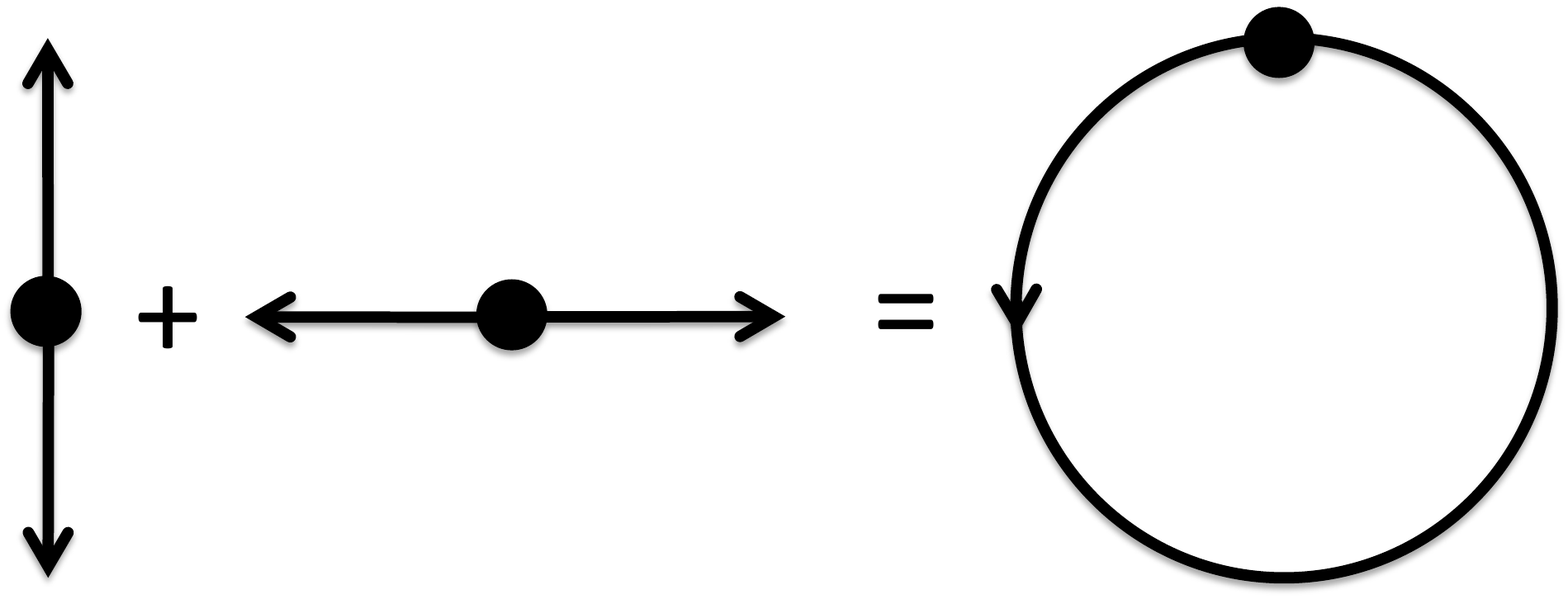}
\vspace*{-1cm}\caption{The two linear zero angular momentum modes of a two dimensional oscillator and
the circular maximal angular momentum ("yrast")mode. }
\label{f:osci}
\end{minipage}
\hspace{0.5cm}
\begin{minipage}[b]{0.47\linewidth}
\centering
\includegraphics[width=\columnwidth]{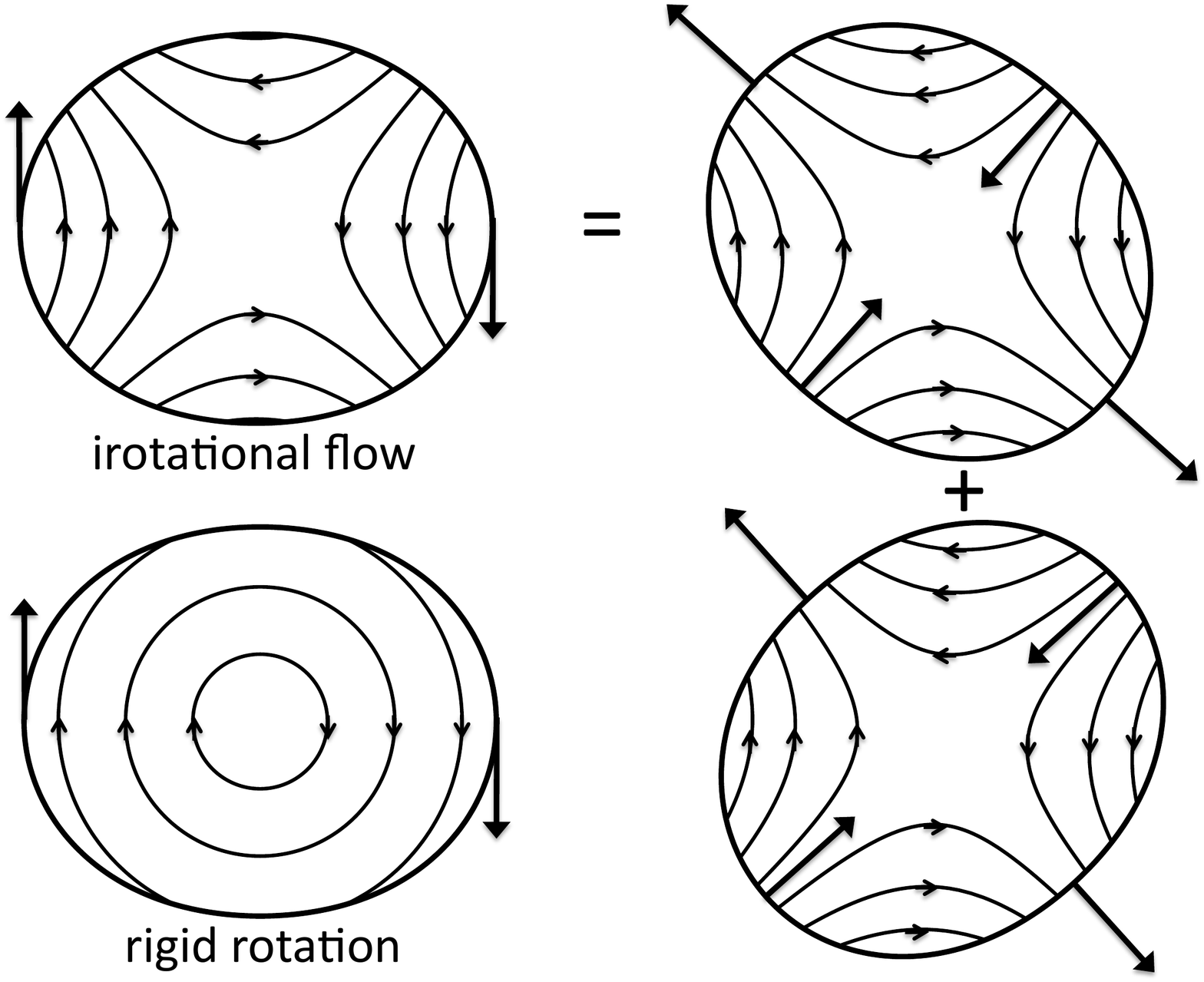}
\caption{Flow pattern of a wave running  about the surface of a drop of ideal
liquid (upper left), standing waves  (pulsating vibrations) of such a droplet (right),  and a rigidly rotating body (lower left).}
\label{f:flow}       
\end{minipage}
\end{figure}

In order to elucidate  the concept,
consider a body of mass $D$ moving in two dimensions subject to a central force. If the force is generated by a (massless) wire
of length $r$, the mass moves on a circle. The angular momentum 
$J={\cal J}\omega$
 increases linearly with the angular velocity $\omega$ while the moment of inertia 
${\cal J}=Dr^2$ is constant.
The energy 
$ E=\omega^2{\cal J}/2=J^2/(2{\cal J})$ is quadratic in $\omega$ or $J$.
 One would classify the system as a rotor. Now consider the vibrator  shown in Fig. \ref{f:osci}.
 The mass is subject to the force $\vec F=-C\vec x$.
The two perpendicular linear normal modes $x_{1,2}=r_{1,2}\cos(\Omega t+\delta_{1,2})$ have the same frequency 
$\Omega=\sqrt{C/D}$. Combining them with the same amplitude $r$ and different phases generates the 
 family of elliptical orbits with the same  energy $Cr^2$, which corresponds to a vibrational multiplet in the quantal context.
 If the phase difference $\delta_1-\delta_2=0,\pi$, the orbits are  linear oscillations carrying no angular momentum. 
 If  $\delta_1-\delta_2=\pm \pi/2$, the orbits  are circles with radius $r$. They carry the maximal angular momentum $J=Dr^2 \Omega$ for the  given energy of $E=Cr^2=\Omega J$, that is, they are  "yrast".   One may view the yrast orbits in an alternative way. The mass moves with the angular velocity $\Omega$ on a
 circular  orbit. As in the case of the rotor, the centrifugal force $\vec x D \Omega^2$ is balanced by the 
 centripedal force $\vec F$. In the case of  $\vec F=-\vec x C$
 this is   only possible  if  $\Omega=\sqrt{C/D}$, which is independent of $r$. 
 That is, the vibrator and the rotor have the same yrast orbits, the circular motion. The difference is the relation between 
 $E,~J,$ and $ \omega$.  In the case of the rotor, the moment of inertia ${\cal J}=Dr^2$ is constant. $E$ and $J$ increase
  due to the  increase $\omega$, resulting in the quadratic relation $E=J^2/2{\cal J}$. In the case of the vibrator 
  $\omega=\Omega$ is constant.
 $E$ and $J$ increase due to an increase of ${\cal J}$ (or equivalently of the radius $r$), resulting in the linear relation
 $E=\Omega J$. In the general case of a centripedal force $\vec F=-\nabla V(|\vec x|)$ the yrast orbit is
 also a circle. Its radius $r_0$ is found by minimizing the energy $E(r)=J^2/2{\cal J}(r)+V(r)$ with respect to $r$
 at fixed angular momentum $J$.  

The case of nuclei is completely analogous. Consider the classical motion of the collective quadrupole modes of the nuclear surface described in the framework of  the phenomenological Bohr Hamiltionian  \cite{BMII}. 
 Solving the classical equations of motion for the deformation parameters $\alpha_\mu(t)$,  one finds that 
the solution with maximal angular momentum 
for given energy (yrast mode) is a uniform rotation about the axis (z-) with
the largest moment of inertia, and static deformation parameters $\beta_e$ and $\gamma_e$.
\begin{eqnarray}  \label{tidalclass}
R(t)= 
R_0[1+\beta_e\cos\gamma_e Y_{20}(\vartheta)
+\sqrt{2}\beta_e\sin\gamma_e\cos(2\varphi-2\omega t)Y_{22}(\vartheta,\varphi=0)].
\end{eqnarray}  
The deformation parameters
$\beta_e,\gamma_e$ are the equilibrium values corresponding to the minimum of the energy
\begin{eqnarray}\label{Ebeta}
E(J,\beta,\gamma)=\frac{J^2}{2{\cal J}(\beta,\gamma)}+V(\beta,\gamma),~~
{\cal J}=4D\beta^2\sin^2(\gamma)
\end{eqnarray}
at given angular momentum $J={\cal J}_3\omega$, where $\omega=dE/dJ$. 

In the case of harmonic vibrator $V=C\beta^2/2$.
Minimizing the energy (\ref{Ebeta}), one finds
$\gamma_e=\pi/2,~ \beta^2_e=J/(2\sqrt{DC})$. The energy
becomes $ E=\omega J=\Omega J/2.$ 
The wave travels with an angular velocity being one half of the oscillator frequency
$\Omega$.  The angular momentum grows due to the
increase of $\beta_e$ which increases moment of inertia ${\cal J}=4D\beta^2_e=J/(2\sqrt{DC})$.
The increase of  deformation is reflected by the reduced transition probability
$B(E2)\propto\beta_e^2\propto J$.
We suggest calling this mode a "tidal wave" because it looks like
the tidal waves  on the 
oceans of our planet. \mbox{Fig. \ref{f:flow}} upper left illustrates the motion of the 
surface and shows the stream lines. 
The right part shows that  the running wave of the five-dimensional quadrupole oscillator 
can be seen as the combination of two standing waves pulsating with a phase
 shift of $\pi/2$. The standing waves carry  no angular momentum. In the case of the
quantized oscillator, the $n$ -phonon states form a multiplet.
Within the multiplet, the tidal wave represents the state with maximal angular momentum 
$I=2n$ and the standing wave the state with zero angular momentum $I=0$.  
The latter is usually invoked in the context of a vibrator (V) - like nucleus. 

The analogy with real tidal waves has the caveat that water is not an ideal liquid. 
 The oceans carry damped surface waves, which in their stationary state
 oscillate with the frequency of the  driving gravitational force. The surface wave on the droplet of 
ideal nuclear liquid runs with an angular velocity corresponding  to the natural  
frequency of the undamped vibrator.   

The uniform rotation of a
statically deformed shape (\ref{tidalclass})  is the yrast mode for any potential $V(\beta,\gamma)$. If 
$V$ has a pronounced minimum at $\beta_0$ and $\gamma_0$
 then $\beta_e\approx\beta_0$ and $\gamma_e\approx\gamma_0$. The moment
 of inertia ${\cal J}\approx const.$ and $J\propto \omega$. We refer to this case as a rotor (R).
 We classify the yrast states of V- like  nuclei as "tidal waves": The deformation
 strongly increases with $J$ while $\omega$ stays nearly constant. For the
 R-like nuclei the deformation stays nearly constant while $J\propto\omega$.
 The transition between the V and R regimes is gradual. 

The left part of Fig. \ref{f:flow} illustrates that same uniform rotation of the surface  may be 
generated in different ways, by the irrotational flow of an ideal liquid or by rigid rotation. 
Yet, the flow pattern of real nuclei differs from both of these ideal cases. It must be calculated in a microscopic way, which   
can be achieved by finding the selfconsistent  mean field solution in a 
rotating frame of reference.  We use the cranking model 
for  calculating the tidal wave modes in vibrational and transitional nuclei.
Marshalek \cite{marshalek} demonstrated that the RPA equations 
for harmonic quadrupole vibrations of nuclei with a spherical 
ground state are a
solution to the selfconsistent cranking model if 
the deformation of the mean field is treated as a small perturbation.
In this paper we solve numerically the cranking problem without the small deformation
approximation, which allows us to describe the yrast states of all 
nuclei in the range between  the harmonic vibrators  and rigid rotors.
Such microscopic approach takes into account
not only the quadrupole shape degrees of freedom, 
as in the above discussed phenomenological model, but also the
the quasiparticle degrees of freedom, which play an important role to
 be discussed below. Of course, the cranking model also applies to the rotational nuclei.
 
\begin{figure}[t]
\vspace{-1cm}
\center
\includegraphics[width=10cm]{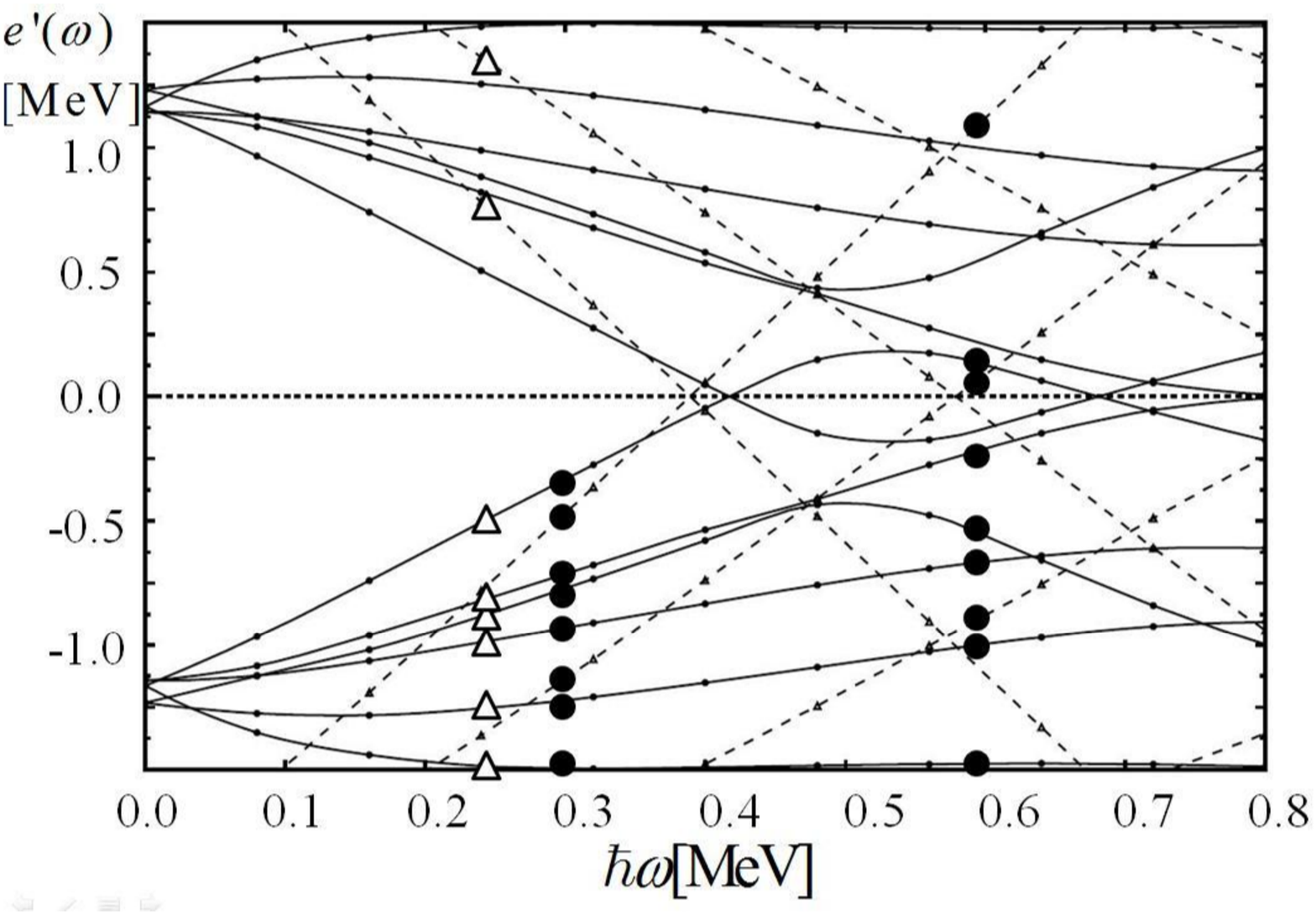}
\vspace{-0.5cm}
\caption{The Quasi Neutron Routhians for $N=56$ and \mbox{$\varepsilon_2=0.13$, $\gamma=0$}.
Full lines: positive parity, dash dotted lines: negative parity (h$_{11/2}$). The full circles show the g-configuration,
which corresponds to a tidal wave, the open triangles show the s-configuration.}
\label{f:qneutron}       
\end{figure}

We use
the Shell Correction version of the 
Tilted Axis Cranking model (SCTAC) 
in order to study the tidal waves in the nuclides with 
$Z=$44, 46, 48 and $N=$56, 58, 60, 62, 64, 66. 
The details and parameters of SCTAC are described in Ref.\cite{qptac}. 
For the yrast states of even-even nuclei, 
the axis of rotation coincides with a principal axis (x). 
The condition $<J_x>=J=I$ 
is used to fix $\omega$, which is solved numerically for each point on
a grid in the $\varepsilon_2-\gamma$ plane. In solving
it is crucial to employ the diabatic tracing technique described in  Ref.\cite{qptac}, 
which is illustrated in Fig. \ref{f:qneutron}.

\begin{figure}[t]
\center
\includegraphics[width=14cm]{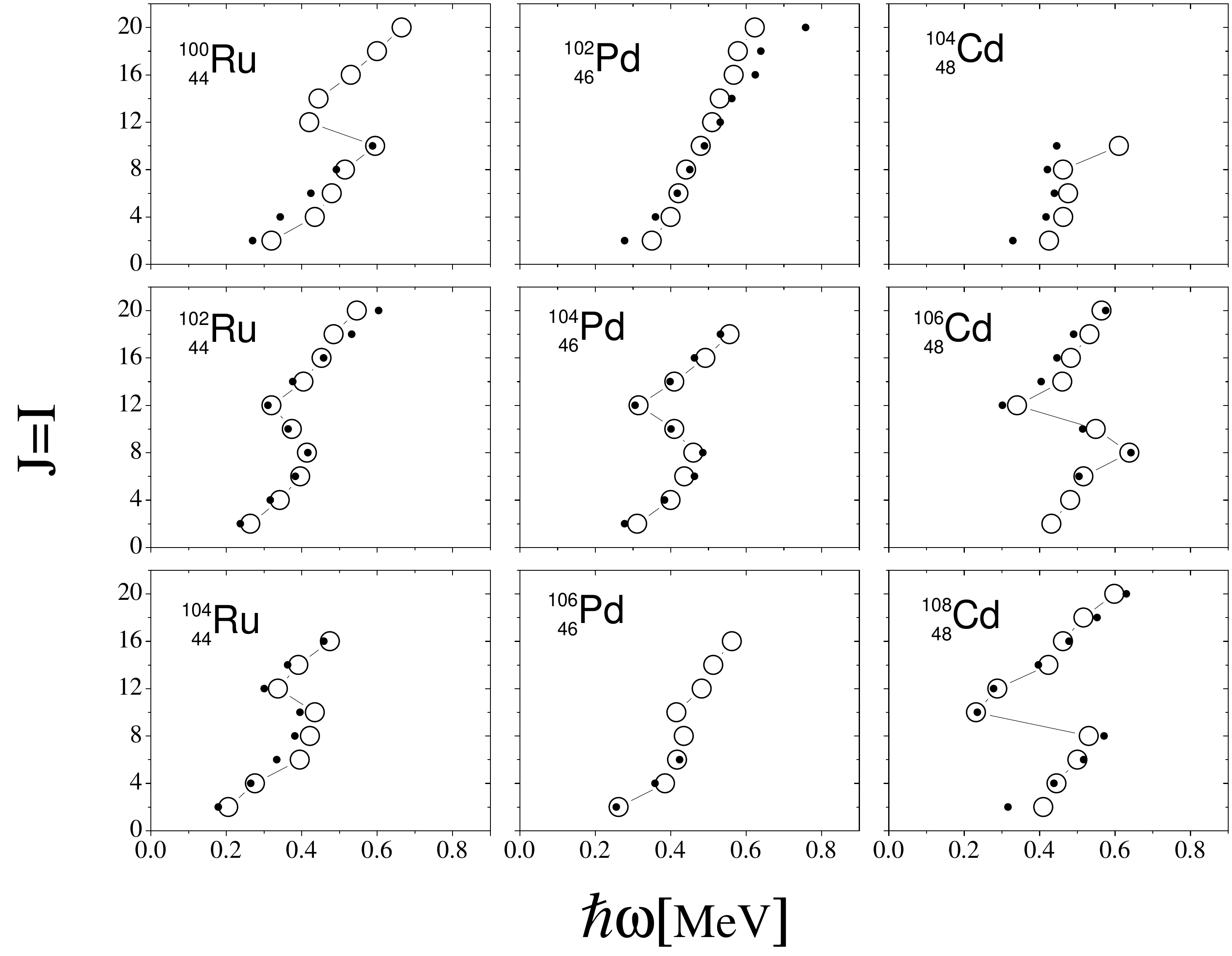}
\includegraphics[width=14cm]{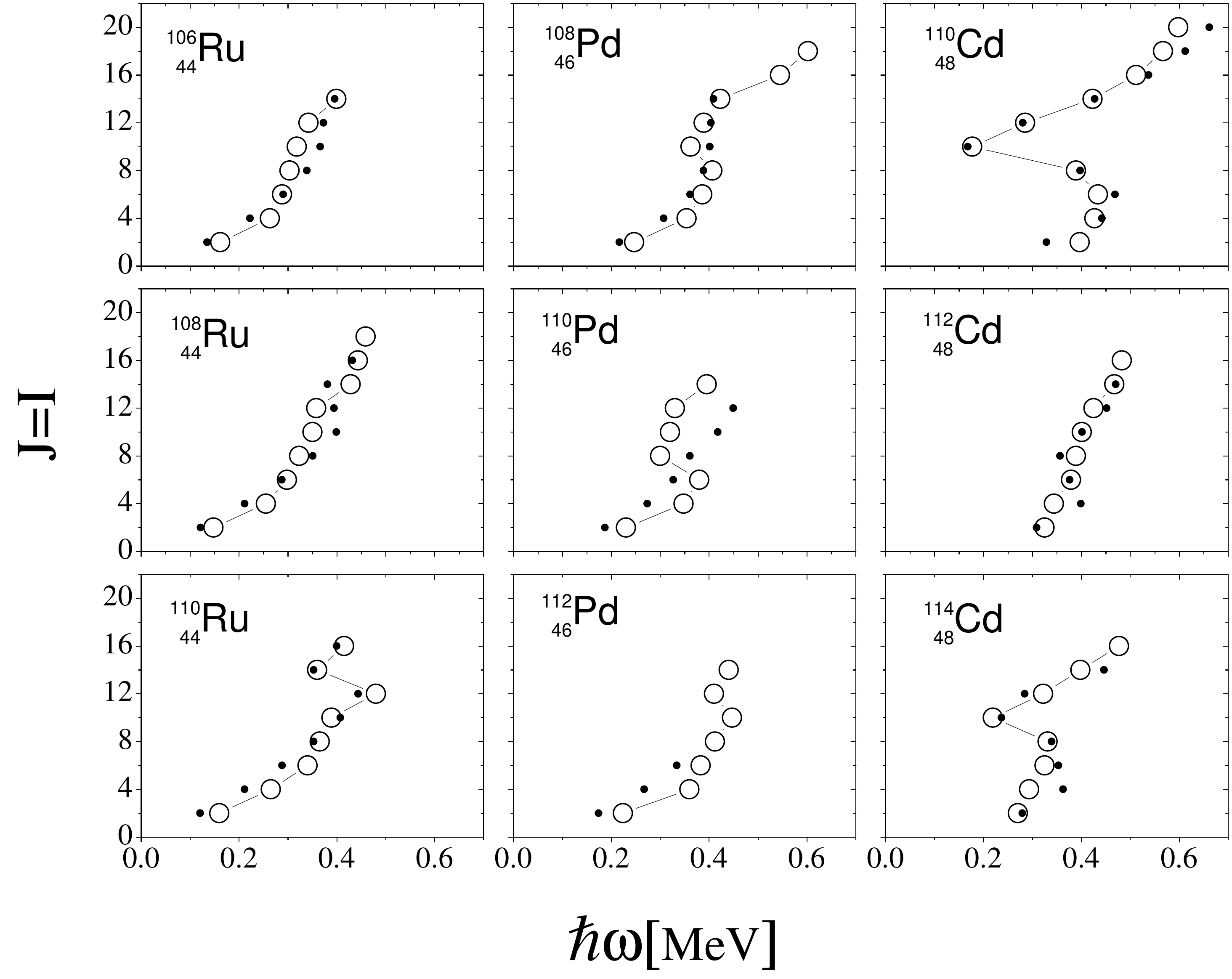}
\caption{Angular momentum $J$ as function of the angular frequency
$\omega$ in the mass 110 region.  
 Full circles show the experiment, open the calculations.
 The experimental functions $J(\omega)$ and ${\cal J}(I)$
 are defined in the standard way by as 
$ \omega(I)=(E(I)-E(I-2))/2,$
$J(I)=I-1/2$,
${\cal J}(I)=J(I)/\omega(I)$. 
Data from\protect \cite{ensdf}.}
\label{f:tidaliom}       
\end{figure}

\begin{figure}[t]
\center
\includegraphics[width=14cm]{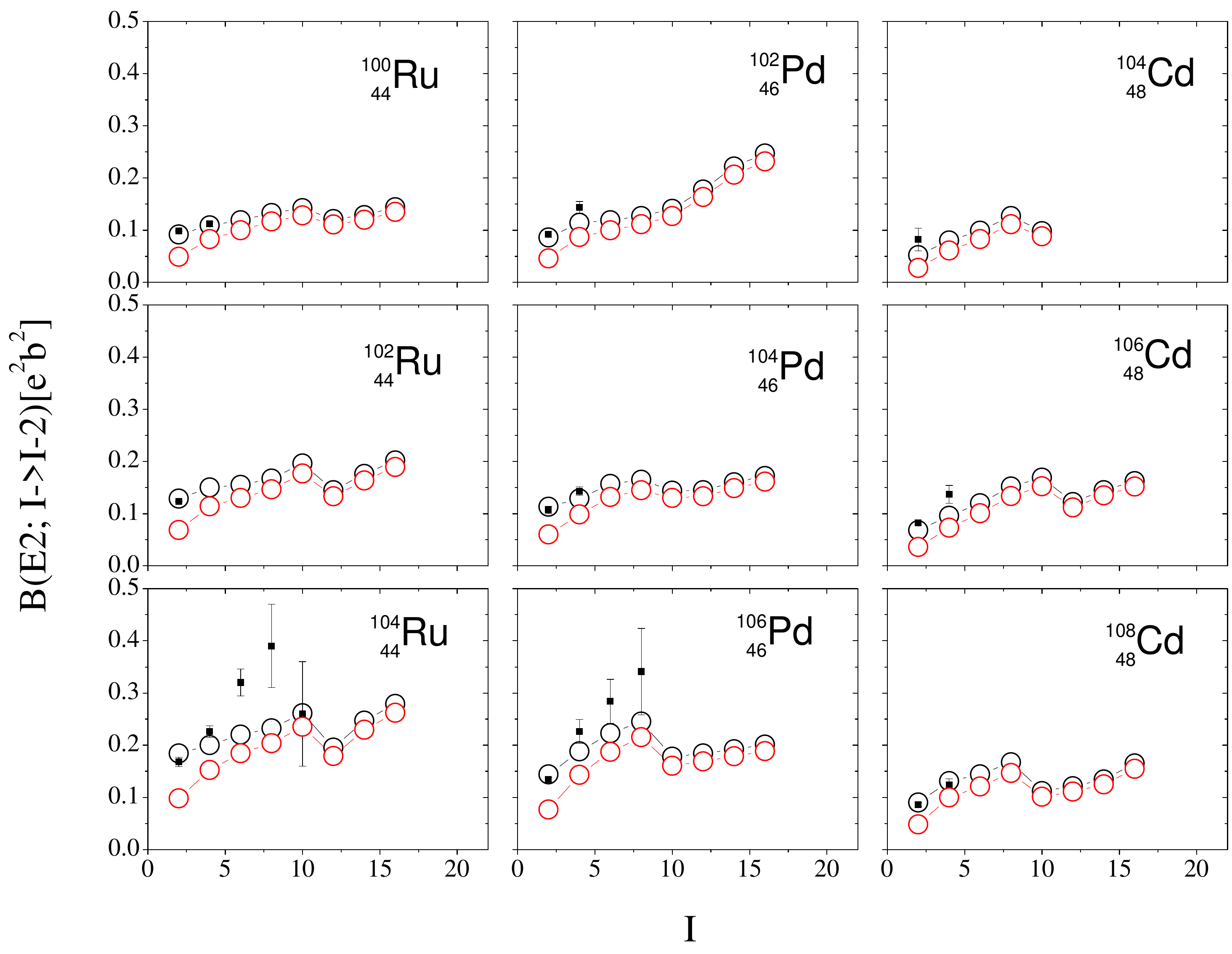}
\includegraphics[width=14cm]{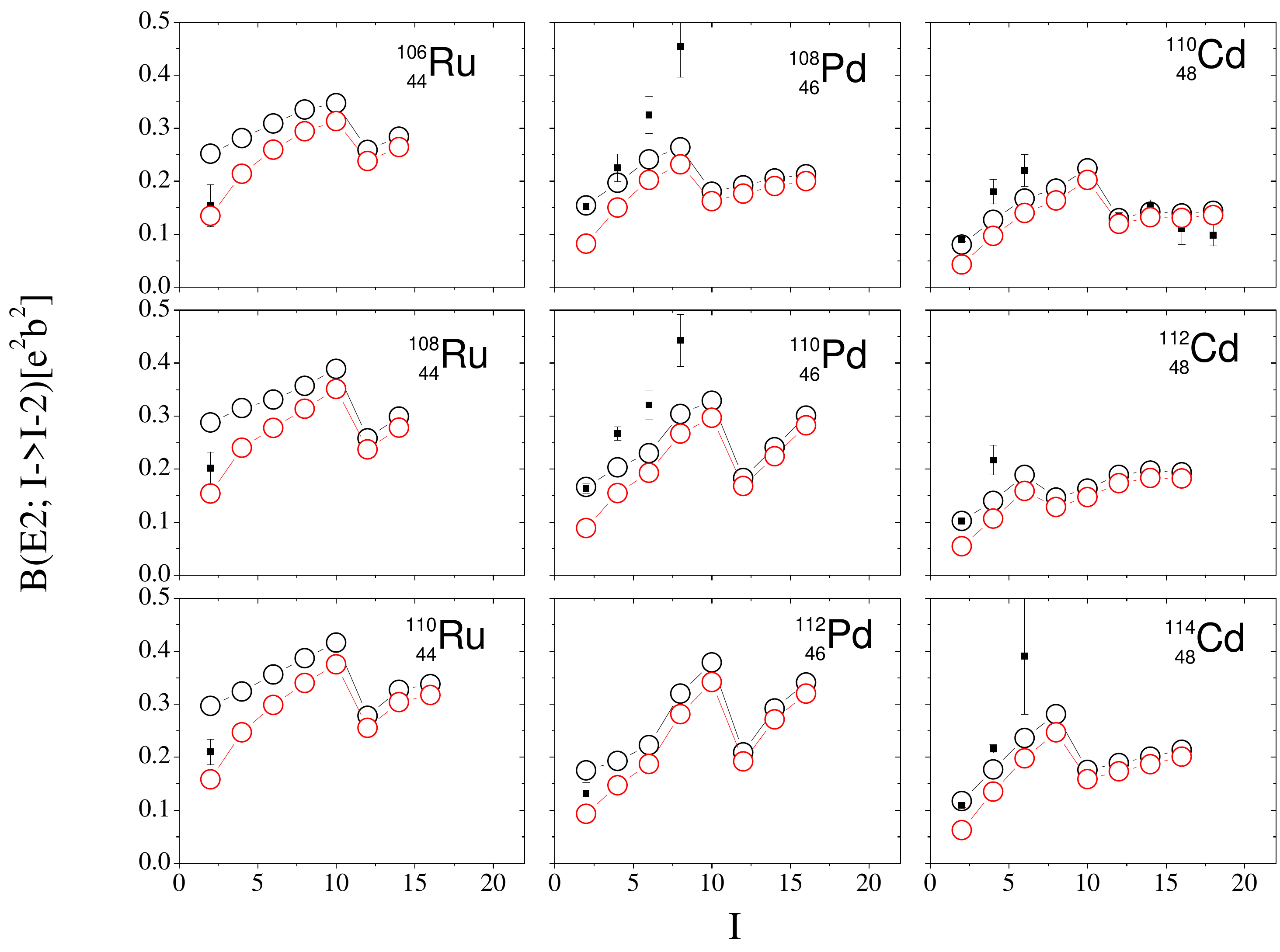}
\caption{The $B(E2, I->I-2)$ values in the mass 110 region.
Full squares show the experiment, open circles the calculations. The higher circles (black)
are the high-spin limit, the lower circles (red) include the low-spin correction
factor (see text). The $B(E2)$ values lie between the two values, where
V-like nuclei are expected closer to the higher values  and R-like nuclei
closer to the lower values.  Data from \protect
\cite{raman,ensdf}.}
\label{f:tidalbe2}       
\end{figure}

The interpolated function $E(\varepsilon_2,\gamma,J)$ 
is minimized with respect to the deformation parameters for 
fixed angular momentum. The final 
value of $\omega(J)$ for the equilibrium deformation is
found by interpolation between the values at the grid points.
It is shown as the theoretical points in the Figure \ref{f:tidaliom}.
It is noted, that the standard technique of finding the selfconsistent
solutions by iteration at fixed $\omega$ becomes problematic
for tidal waves close to the harmonic limit because the total Routhian is 
nearly deformation independent.           
The pairing parameters  are kept fixed to $\Delta_n=\Delta_p=1.1$ MeV, 
and the chemical potentials are $\lambda$ adjusted to the particle numbers at $\omega=0$. 
We slightly correct
\mbox{$
 J(\omega)=J_{SCTAC}(\omega)+ 100MeV^{-1} \varepsilon_2^2\sin^2(\gamma-\pi/3)\omega,$}
which adds less than 0.5 $\hbar$ to the angular momentum. About half of 
it takes into account the coupling between the oscillator
shells  and another half is expected
to come from quadrupole pairing, both being neglected in SCTAC. 
The changes due to the correction are smaller than the radius of the
open circles in Figs. \ref{f:tidaliom} and \ref{f:tidalbe2}.  

The $B(E2,I->I-2)$ values are calculated for $J=I$ as
described in \cite{qptac} for $\theta=90^o$, which is the high-spin limit.
In addition we give the values multiplied by the factor 
$I(I-1)/(I+1/2)(I-1/2)$, which corrects for the low-spin
deviation of the $B(E2)$ values from the high-spin limit in the case
of an axial rotor. For vibrators, the
correction factor should not be applied \cite{marshalek}, for rotors it should.
The correction factor accounts for zero point fluctuations of the angular momentum
of a rigid axial rotor.
Since the present approach does not take into account the zero point motion 
of the collective variables in a systematic way, one should consider the two values 
in Fig. \ref{f:tidalbe2} as limits, between which the $B(E2)$ values lie.  

\begin{figure*}[h]
\begin{minipage}[b]{5cm}
\center
\includegraphics[width=\linewidth]{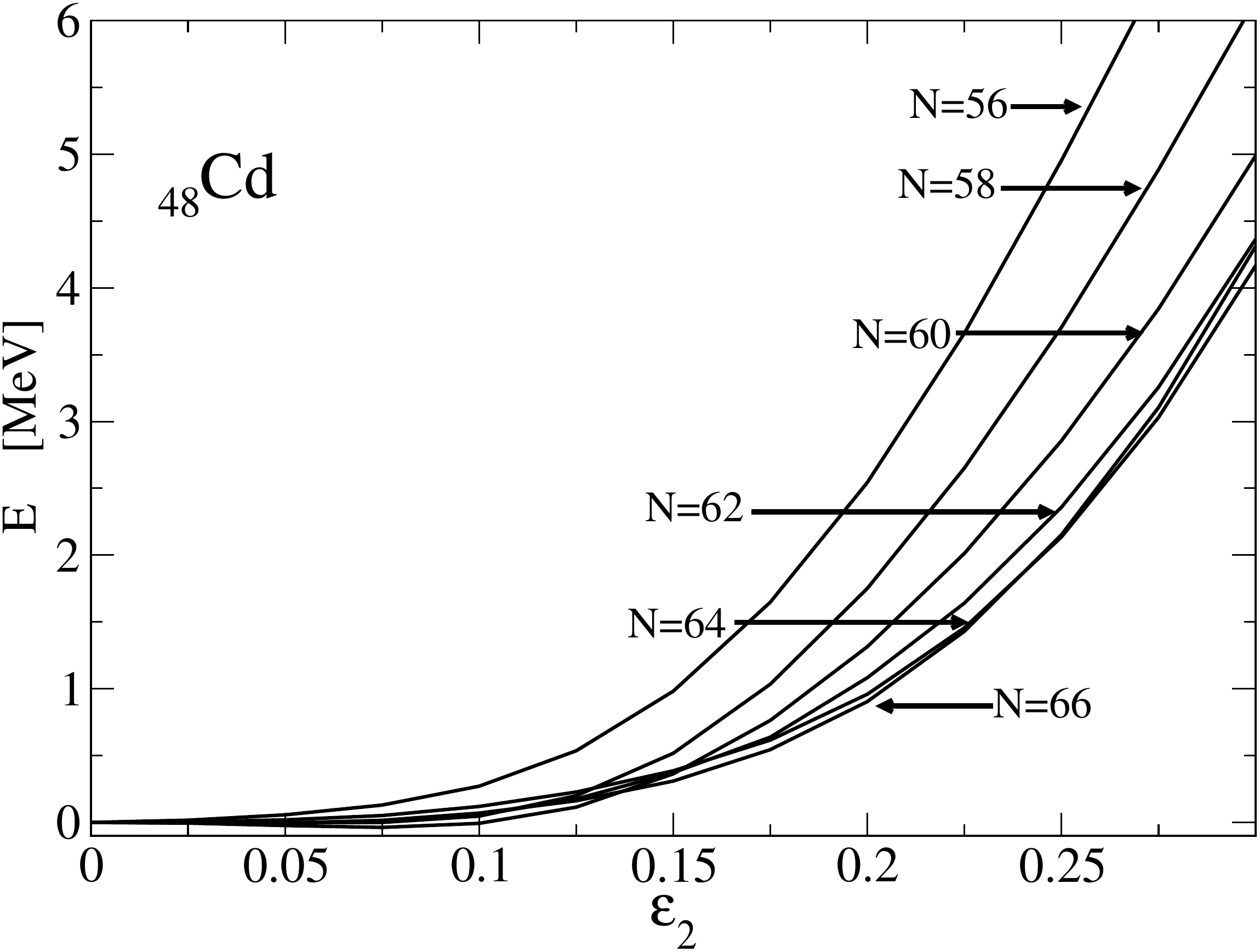}
\end{minipage}
\hfill \mbox{} \hfill
\begin{minipage}[b]{5cm}
\center
\includegraphics[width=\linewidth]{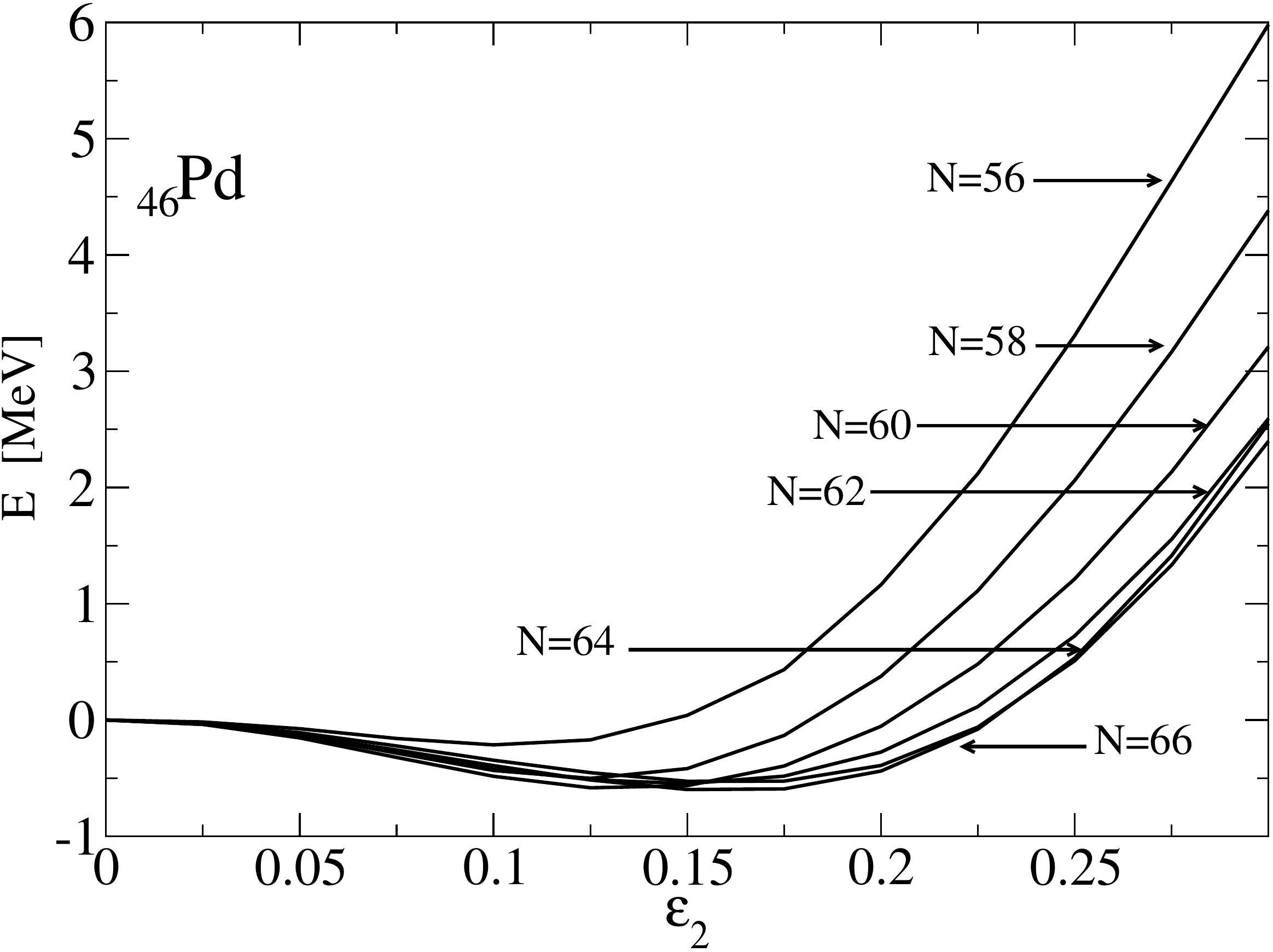}
\end{minipage}
\hfill \mbox{} \hfill
\begin{minipage}[b]{5cm}
\center
\includegraphics[width=\linewidth]{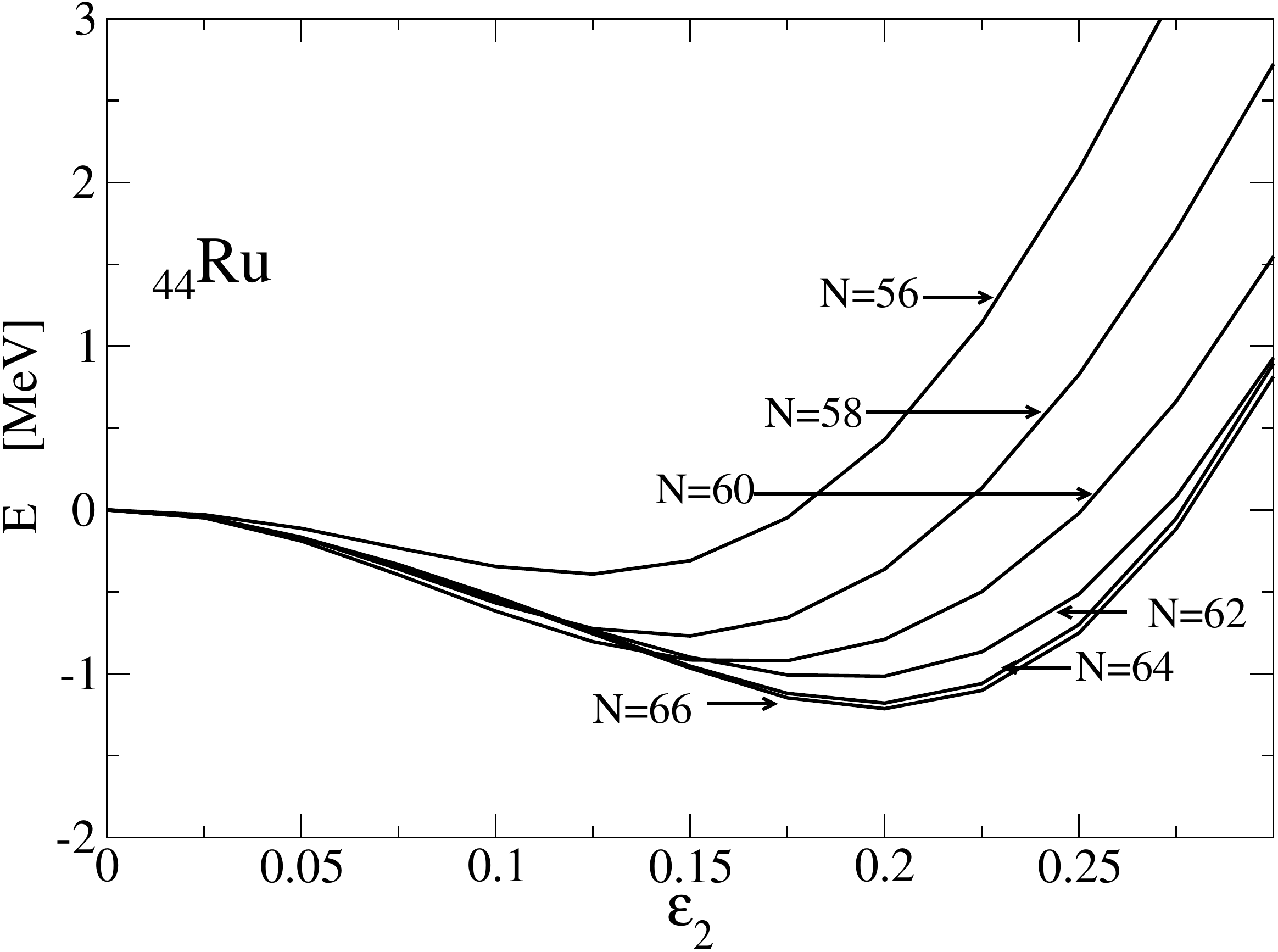}
\end{minipage}
\caption{\label{f:gsenergy}
The ground state energies at $\gamma=5^\circ$. }
\end{figure*}

The ground state energies are shown  in Fig. \ref{f:gsenergy}. They are 
calculated using the modified oscillator potential with the parameters
given in Ref. \cite{qptac}, adding a small adjustment $k \varepsilon_2^2$ with 
$k=$8.5, 7.9, 8.5, 9.6, 11.9, 14.5 MeV for $N$= 56, 58, 60, 62, 64, 66, respectively.
The correction term slightly disfavors deformation.  It is
substantially smaller  than the differences between deformation energies 
calculated from the various mean field theories used at present.
The tendency to deformation increases with the numbers of 
valence proton holes and valence neutron particles in the $Z=N=50$ 
shell. The $Z=48$ isotopes are spherical, becoming softer 
with increasing $N$.   
The $Z=44$ isotopes are slightly deformed, very soft, becoming less
soft with $N$. The $Z=46$ isotopes are intermediate. The equilibrium deformations 
for $2\leq I$ are $\gamma_e=0^\circ$ and $0.1<\beta_e<0.2$, except
$^{110,112}$Cd for which $\gamma=10^\circ$ - $15^\circ$ for $2\leq I\leq 8$. 
The heavy Pd isotopes   have a larger $\beta_e$. 
Figs. \ref{f:tidaliom} and \ref{f:tidalbe2} compare
the calculated values of $\omega(J)$ and $B(E2, I->I-2)$ with experiment.
The calculated frequencies reproduce the experimental ones very well.
The calculated $B(E2)$ in the V-like nuclei increase somewhat slower with
$I$ than in experiment.
The back-bend in the $J(\omega)$ curves is caused by the rotational  
alignment of a pair of $h_{11/2}$ neutrons (change from the g-configuration 
to the s- configuration in Fig. \ref{f:qneutron}). 
The sharpness of the alignment depends
sensitively on the position of the neutron chemical potential $\lambda_n$ 
relative to the  $h_{11/2}$ levels \cite{Hamamoto}, which explains why
there is a strong back-bend for some $N$ and only a smooth up-bend for other $N$.
The $h_{11/2}$ levels  move with the changing deformation, and
differences of the deformation are the reason
why $J(\omega)$ sharply back-bends in  $Z=48,~N=62$ while it gradually grows
in $Z=44,~N=62$. Also normal parity neutron orbitals contribute to the $J(\omega)$
in a non-linear way, in particular at the avoided crossings  between them 
(around $\hbar \omega = $ 0.53 MeV in Fig. \ref{f:qneutron}).
 For example, they are responsible that in $^{110}$Cd
 the  point $I=8$ has a lower $\omega$ than the point $I=6$. 
  The large frequency
encountered at small deformation makes also other orbitals than the high-j intruders react
in a non-linear way to the inertial forces.  For many nuclides 
the quasiparticle degrees cannot be treated in a perturbative way for $I \geq 6$, which means
the separation of the collective quadrupole degrees of freedom becomes
problematic. For the nuclei with the smallest numbers of valence particles
and holes this entanglement of collective and single particle degrees appears
already for $I=4$, which is reflected by the irregular curves $J(\omega)$ for
$Z=48,~N=56$  in Fig. \ref{f:tidaliom}.

As seen in Fig. \ref{f:tidalbe2},
the aligned $h_{11/2}$ quasi neutrons
 stabilize the deformation at a {\it smaller}
 value than reached before, which is reflected by the decrease 
of the $B(E2)$ values. The deformation increases slower than before the
alignment, which means the motion becomes more rotational.
In some cases, the alignment process is extended over several
units of angular momentum, keeping $\omega$ nearly constant.
This means the simple picture of tidal waves, as running at nearly
 constant angular velocity and gaining angular momentum by 
deforming does not account for the 
full complexity of real nuclei. The single particle degrees of freedom also
provide their share of  angular momentum, such that the rotational frequency remains
constant.

$^{102}$Pd is a special case.  The g- and s-configurations do not mix, and are both
 observed up to $I=20$. Figs. \ref{f:tidaliom} and \ref{f:tidalbe2} show only the 
 g-sequence, which represents a tidal wave that gains angular momentum 
 predominantly by increasing the deformation, as discussed above for the
 droplet model. Its $I=16$ state corresponds to eight stacked phonons! 
 Only for   $I >16$ the alignment of the positive parity neutrons 
 (avoided crossing at \mbox{$\hbar \omega \approx 0.53$ MeV} in Fig. \ref{f:qneutron})
 becomes a comparable source of angular momentum.

In conclusion, the yrast states of vibrational
and transitional nuclei can be understood as tidal waves that run over the nuclear 
surface, which have a static deformation, the wave amplitude, in the frame of 
reference co-rotating
with the wave. In contrast to rotors,  most of the angular momentum is 
gained by an increase of the deformation  while the angular velocity
increases only weakly. Since tidal waves have a static deformation in the rotating frame,
the rotating mean field approximation (Cranking model) provides a microscopic
description. The yrast states 
of  ``spherical'' or weakly deformed nuclei with $44\leq Z\leq 48,~56\leq N\leq66$ were calculated up to
spin 16. These microscopic calculations reproduce the energies and 
electromagnetic $E2$ transition probabilities in detail.
 The structure of the yrast states turns out to be complex. The shape
degrees of freedom and the single particle degrees of freedom are intimately
 interwoven. The non-linear response to the inertial forces
of the individual orbitals at the Fermi surface determines the
way the angular momentum is generated. This lack of separation 
of the collective motion from the single particle becomes increasingly important 
for increasing angular momentum and decreasing number of valence nucleons. 

Supported by the DoE Grant DE-FG02-95ER4093.


\begin{thebibliography}{}


\bibitem{arXiv} S. Frauendorf, Y. Gu, and J. Sun, arXiv-id: 0709.0254 (2007)
\bibitem{BMII} A. Bohr and B. R. Mottelson, 
Nuclear Structure, Vol. II (Benjamin, New York, 1975).
\bibitem{ensdf} ENSDF data bank, http://www.nndc.bnl.gov
\bibitem{marshalek}E.R. Marshalek, Phys. Rev. C {\bf 54} 159 (1996). 
\bibitem{qptac}S. Frauendorf, Nucl. Phys. A {\bf 677}, 115 (2000). 
\bibitem{raman}S. Raman {\it et al.}, Atomic D. and Nucl. D. Tab. {\bf 78}, 1 (2001)
\bibitem{Hamamoto} R. Bengtsson, {\it et al.}, Phys. Lett. B  {\bf 73},  259 (1978) 
\end{thebibliography}
\end{document}